\documentclass[conference, 10pt]{IEEEtran}
\IEEEoverridecommandlockouts
\usepackage{cite}
\usepackage{mathrsfs, amsmath,amssymb,amsfonts}
\usepackage{graphicx}
\usepackage{textcomp}
\usepackage[table]{xcolor}
\usepackage{nicefrac}

\def\BibTeX{{\rm B\kern-.05em{\sc i\kern-.025em b}\kern-.08em
    T\kern-.1667em\lower.7ex\hbox{E}\kern-.125emX}}

\usepackage{lipsum}

\usepackage[utf8]{inputenc}
\usepackage[T1]{fontenc}

\usepackage[abbreviations]{foreign}
\usepackage{algorithm}
\usepackage{algpseudocode}

\usepackage{xspace}

\usepackage{listings}
\newcommand{\code}[1]{\texttt{\small #1}\xspace}

\usepackage{subcaption}
\usepackage{textcomp}
\usepackage{booktabs}

\usepackage{tabularx}

\usepackage{url}
\usepackage{pifont}
\usepackage[colorlinks=true,linkcolor=blue,anchorcolor=blue,citecolor=blue,filecolor=black,menucolor=black,runcolor=black,urlcolor=blue]{hyperref}

\usepackage[inline]{enumitem}
\setlist{noitemsep,topsep=0pt,parsep=0pt,partopsep=0pt}

\usepackage[capitalise]{cleveref}

\usepackage{tikz}
\usetikzlibrary{positioning, arrows.meta}
\usetikzlibrary{matrix}

\definecolor{lightcolor}{rgb}{0,0.5,1}

\usepackage{fontawesome5}
\newboolean{showcomments}
\setboolean{showcomments}{true}
\ifthenelse{\boolean{showcomments}}
{ \newcommand{\mynote}[3]{
		\fbox{\bfseries\sffamily\scriptsize#1}
		{\small$\blacktriangleright$\textsf{\emph{\color{#3}{#2}}}$\blacktriangleleft$}}}
{ \newcommand{\mynote}[3]{}}

\definecolor{darkgreen}{rgb}{0.3,0.5,0.3}
\definecolor{darkblue}{rgb}{0.3,0.3,0.5}
\definecolor{darkred}{rgb}{0.5,0.3,0.3}

\newcommand{\dpuhost}{\texttt{dpu-host}\xspace}
\newcommand{\hostdpu}{\texttt{host-dpu}\xspace}
\newcommand{\cpu}{\texttt{cpu-time}\xspace}
\newcommand{\dpu}{\texttt{dpu-time}\xspace}

\newcounter{numobserv} 
\definecolor{beaublue}{rgb}{0.88, 0.93, 0.93}
\usepackage{tcolorbox}
\colorlet{shadecolor}{beaublue}
\tcbset{
	colback=beaublue,
	boxrule=0.5pt,
	boxsep=0pt,
	left=1pt,
	right=1pt,
	top=1pt,
	bottom=1pt,
	sharp corners,
	colframe=white,
}
\newcommand{\observ}[1]{
	\addtocounter{numobserv}{1}
	\begin{tcolorbox}	
		\textit{\textbf{Take-away\,\thenumobserv\,:} #1 }	
	\end{tcolorbox}
}

\begin{document}


\title{Evaluating the Potential of In-Memory Processing to Accelerate Homomorphic Encryption\\
{\large Practical Experience Report}
}

\author{
Mpoki Mwaisela$^\dagger$~~Joel Hari$^\ddagger$~~Peterson Yuhala$^\dagger$~~Jämes Ménétrey$^\dagger$~~ Pascal Felber$^\dagger$~~Valerio Schiavoni$^\dagger$\\
\IEEEauthorblockN{\small$^\dagger$University of Neuchâtel, Switzerland~\{firstname.lastname@unine.ch\}}
\IEEEauthorblockN{\small$^\ddagger$University of Bern, Switzerland~\{firstname.lastname@students.unibe.ch\}}
}

%
%
%
%
%
%
%
%
%
%
%
%


\maketitle

\thispagestyle{plain}
\pagestyle{plain}
%
\begin{abstract}
The widespread adoption of cloud-based solutions introduces privacy and security concerns.	
Techniques such as homomorphic encryption (HE) mitigate this problem by allowing computation over encrypted data without the need for decryption.
However, the high computational and memory overhead associated with the underlying cryptographic operations has hindered the practicality of HE-based solutions. 
While a significant amount of research has focused on reducing computational overhead by utilizing hardware accelerators like GPUs and FPGAs, there has been relatively little emphasis on addressing HE memory overhead. 
Processing in-memory (PIM) presents a promising solution to this problem by bringing computation closer to data, thereby reducing the overhead resulting from processor-memory data movements.
In this work, we evaluate the potential of a PIM architecture from UPMEM for accelerating HE operations.
Firstly, we focus on PIM-based acceleration for polynomial operations, which underpin HE algorithms.
Subsequently, we conduct a case study analysis by integrating PIM into two popular and open-source HE libraries, OpenFHE and HElib.
Our study concludes with key findings and takeaways gained from the practical application of HE operations using PIM, providing valuable insights for those interested in adopting this technology.
\end{abstract}


\section{Introduction}
\label{sec:intro}
The increased reliance on cloud computing for data processing presents critical challenges with regard to security and privacy.
This is because potentially sensitive data is entrusted to untrusted cloud service providers.
Moreover, recent security and privacy regulations, \eg the EU's General Data Protection Regulation (GDPR)~\cite{GDPR2016}, the California Consumer Privacy Act (CCPA)~\cite{CCPA2018}, the Brazilian General Data Protection Act (LGPD)~\cite{LGPD2018}, require the cloud providers to adopt data processing techniques that preserve the privacy of the data owners.
However, standard data encryption techniques require the ciphertext to be decrypted prior to being processed, hence defeating the goal of guaranteeing privacy.

In contrast, \emph{homomorphic encryption} (HE) allows computation over encrypted data, preserving privacy by enabling processing without decryption.
A client first encrypts their sensitive data and sends it to an untrusted cloud provider.
The latter performs computations on the encrypted data, \eg bio-medicine~\cite{DBLP:journals/csur/WoodNK20}, genome testing~\cite{DBLP:journals/iacr/LauterLN15, foresee2015}, financial data analysis~\cite{lrfinance2019}, machine learning inference~\cite{gazelle}.
Once the computations are over, the encrypted result is sent back to the client, which can be decrypted in a secure local environment.

In spite of its promising security properties, HE suffers from substantial computational and memory overhead~\cite{feldmann21,saransh21}.
Accelerators like FPGAs and GPUs have been employed to improve the compute throughput of HE algorithms~\cite{cousins17,sinha19,wang12,jung21,10.1145/3620665.3640397}, but these techniques still suffer from memory overheads.
For instance, encrypting an integer in the homomorphic domain increases its size from 4B to over 20KB~\cite{saransh24}.
This rapid growth in ciphertext sizes leads to poor data locality, exacerbating the cost of data movements between the CPU, memory, and potential accelerators. 
Moreover, the limited bandwidth of the memory channel contributes to the \emph{memory wall} or \emph{von Neumann bottleneck}~\cite{nider21}, increasing latency when accessing main memory. 

Processing in-memory (PIM) augments memory with processing capabilities, thereby bringing computation closer to data. 
Our work focuses on UPMEM PIM~\cite{upmem-pim}, the first commercialized PIM architecture~\cite{gomez22}.
This PIM architecture provides conventional DRAM DIMMs comprising general-purpose processors, called \emph{DRAM processing units} (DPUs).
Each 64MB chunk of memory in a UPMEM DIMM is coupled with a single DPU, thus allowing computing capability to scale with memory size~\cite{nider21}. 
This PIM architecture shows promise in addressing the memory overhead encountered in most of the HE algorithms and schemes thanks to its extensive parallelism and improved memory bandwidth.

This practical experience report evaluates the potential of leveraging DPUs to accelerate HE operations, including real-world scenarios using state-of-the-art FHE open-source libraries.
The key contributions of this study are as follows:
\begin{itemize}[leftmargin=*]
    \item Analyzing the performance of polynomial addition and multiplication operations (\ie the building blocks of HE algorithms), across multiple DPUs and assessing the overheads of DPU-based operations and data copy operations. We compare these against single and multithreaded CPU-based baselines.
    \item We implement the first PIM-HE libraries by integrating actual PIM hardware with two widely used open-source HE libraries: OpenFHE~\cite{openfhe22} and HElib~\cite{helib}.
    \item We provide an extensive experimental evaluation of our prototypes, which we release as open source for encouraging further research with PIM.
    \item A thorough discussion on the advantages and drawbacks of using PIM for FHE, grounded in our hands-on experience and results. 
    Additionally, we propose potential solutions to mitigate certain drawbacks identified in our investigations. 
\end{itemize}

\noindent\textbf{Roadmap.} We provide background on PIM, the UPMEM PIM architecture and HE in \S\ref{sec:background}.
Subsequently, in \S\ref{sec:arch}, we explore PIM-based programming, where we outline the overall structure of a DPU-based program, explain how datasets can be split across multiple DPUs, and discuss how existing libraries can be adapted for UPMEM DPUs.
In \S\ref{sec:eval}, we report on the evaluation of generic polynomial operations that underpin HE algorithms, afterwhich we perform a case study analysis of PIM-based HE operations with two widely used HE libraries: OpenFHE~\cite{openfhe22} and IBM's HElib~\cite{helib}.
We discuss the advantages and drawbacks of UPMEM's PIM design in \S\ref{sec:discussion}.
In \S\ref{sec:rw}, we survey prior work on HE and PIM-related architectures.
We conclude and discuss future work in \S\ref{sec:conclusion}.
\section{Background}
\label{sec:background}

\subsection{UPMEM PIM design}

\begin{figure}[!t]
    \centering
	\includegraphics[scale=0.5]{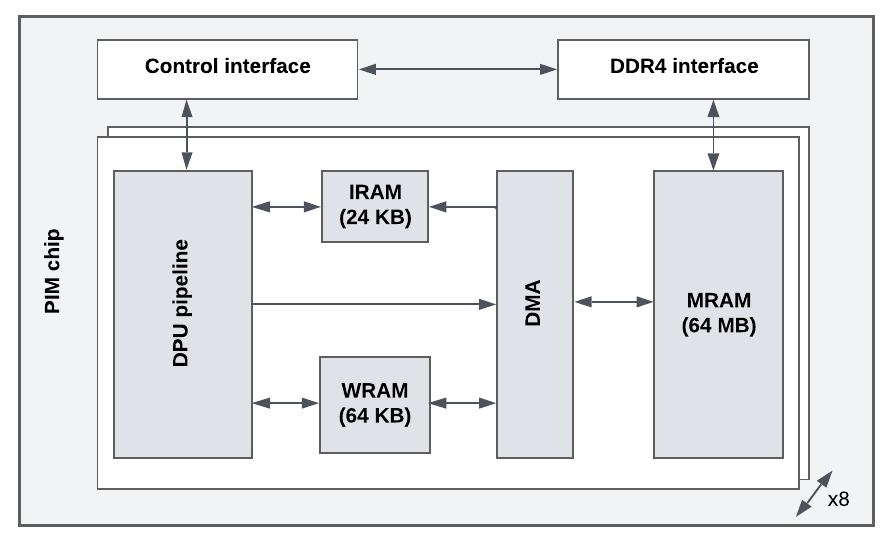}
	\caption{DPU architecture}
	\label{fig:dpu}
	\vspace{-4mm}
\end{figure}
UPMEM PIM modules are standard DDR4-2400 DIMMs, comprising 2 DRAM \emph{ranks}, each containing 8 PIM chips; each chip comprises 8 DPUs, hence 64 DPUs per rank and 128 DPUs per UPMEM DIMM module.
DPUs are general-purpose processors~\cite{nider21} which implement a 32-bit RISC-based instruction set architecture (ISA).
Each DPU shares a 64MB DRAM memory bank with the host CPU called \emph{main RAM} (MRAM).
Each DPU has a 24KB instruction memory called \emph{instruction RAM} (IRAM), and a 64KB scratchpad memory called \emph{working RAM} (WRAM).
Direct communication between DPUs is not supported; all inter-DPU communication must be done via the host CPU, which can transfer data between main memory (DRAM) and MRAM.
UPMEM DPUs follow the \emph{single program multiple data} programming model~\cite{gomez22}. 
Software threads, called \emph{tasklets} which are backed by DPU hardware threads, execute the same program concurrently on different chunks of data.
Each DPU supports up to 24 tasklets.
To prevent memory access conflicts due to concurrent DRAM accesses from the CPU and DPU, the UPMEM PIM architecture maintains distinct address spaces for DPUs and regular DRAM~\cite{lee2024}.
This design imposes the need to perform explicit copy operations of input data from the host DRAM to the DPU and vice versa. 
The DPU ISA has native support for 32-bit integer additions/subtractions and 8-bit integer multiplications.
More complex arithmetic operations, such as integer divisions and floating-point operations, are emulated in software~\cite{rhyner2024AnalysisOD}.

\subsection{Homomorphic Encryption}
\begin{figure}[!t]
\centering
\includegraphics[scale=0.65]{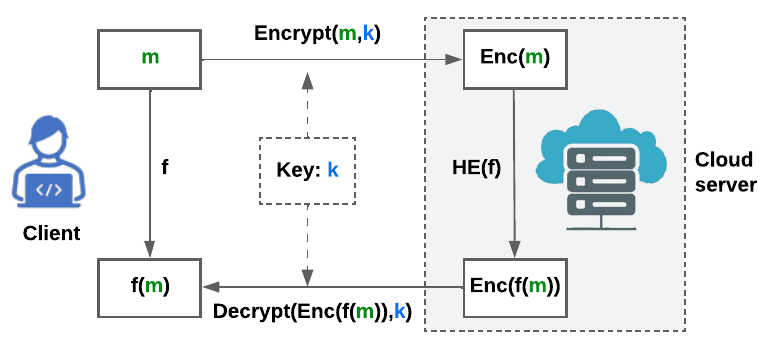}
\caption{Homomorphic encryption allows computation on encrypted data without decrypting it.}
\label{fig:he}
%
\vspace{-2.6mm}
\end{figure}

In cryptography, \emph{encryption} is the process of transforming a (sensitive) piece of data, \ie the \textit{plaintext}, into an alternative form, \ie \text{ciphertext}, which can only be deciphered by authorized entities.
As shown in \autoref{fig:he}, HE allows a computation $f$ to be performed over ciphertexts so that the decrypted result is the same as that obtained from performing the same computation on the plaintexts.
HE can be classified into three categories based on the permitted operations on the encrypted data:
\emph{(1)} \emph{partially homomorphic encryption} (PHE) supports only one type of operation, \eg addition or multiplication, which can be performed an unlimited number of times; 
\emph{(2)} \emph{somewhat homomorphic encryption} (SWHE) supports more than one operation, \eg both addition and multiplication, which can be performed only a limited number of times;
and \emph{(3)} \emph{fully homomorphic encrption} (FHE) allows an unlimited number of operations to be performed an unlimited number of times.
In this study, we focus on FHE, which is the most robust and widely used HE technique. 
Some popular schemes that implement FHE are: Brakerski/Fan-Vercauteren (BFV)~\cite{bfv-1,bfv-2}, Brakerski-Gentry-Vaikuntanathan (BGV)~\cite{bgv2012}, and Cheon-Kim-Kim-Song (CKKS)~\cite{ckks}.
BGV and BFV support only integer arithmetic operations, while CKKS allows computations on complex numbers with limited precision.
Many FHE libraries have been developed to support some or all of these schemes. 
Examples include: OpenFHE~\cite{openfhe22}, IBM's HElib~\cite{helib}, Microsoft SEAL~\cite{sealcrypto}, and Palisade~\cite{palisade}

\smallskip\noindent\textbf{FHE Primitives.} FHE schemes offer a suite of cryptographic primitives that facilitate operations on both plaintext and ciphertext. 
Such primitives include encryption, decryption, and evaluations, \ie \code{EvalAdd} and \code{EvalMult}.
Within the primitives, the plaintext, ciphertext, and keys are encoded as polynomials\cite{helib-opt}, with coefficients sampled over moduli $t$ and $q$, respectively.
As such, the underlying operations performed by a specific FHE primitive translate to mathematical operations over polynomials. \autoref{tab:primitives} outlines some standard HE primitives and the underlying polynomial operations involved.

\renewcommand{\arraystretch}{1.75}
\begin{table*}[ht] 
    \centering
    \caption{Standard cryptographic primitives in HE schemes and the underlying polynomial operations.}
    \begin{tabularx}{\textwidth}{>{\raggedright}X >{\raggedright}p{5.25cm} >{\raggedright\arraybackslash}p{5cm} >{\raggedright\arraybackslash}p{5cm}}
        \toprule
        \rowcolor{gray!10}
        \textbf{Operation} & \textbf{Description} & \textbf{Polynomial Multiplications} & \textbf{Polynomial Additions} \\
        \midrule
        KeyGen & Generates secret key (SK), public key (PK), and evaluation key (EK). & Several, including random polynomial generation and errors. & Minor, mainly adjusting terms. \\[1pt]
        \rowcolor{gray!10}
        Encrypt & Encrypts plaintext into ciphertext using PK. & Multiplication of polynomials in key with noise and message encoding. & Summations of noise and scaled message. \\[1pt]
        Decrypt & Decrypts ciphertext to retrieve plaintext using SK. & Multiplication of ciphertext with SK. & Additions to adjust coefficients and reduce modulo plaintext modulus. \\[1pt]
        \rowcolor{gray!10}
        EvalAdd & Homomorphically adds two ciphertexts. & None directly; operates on existing ciphertexts. & Additions of corresponding polynomial coefficients of ciphertexts. \\[1pt]
        EvalMult & Homomorphically multiplies two ciphertexts. & Multiplication of ciphertext polynomials, including cross terms. & Additions during polynomial product and noise adjustment. \\[1pt]
        \rowcolor{gray!10}
        Relinearize & Reduces the number of polynomial elements in a ciphertext post multiplication. & Multiplications involving EK and ciphertext for dimension reduction. & Summations to combine terms and reduce size. \\
        \bottomrule
    \end{tabularx}
    \label{tab:primitives}
\end{table*}

\smallskip\noindent\textbf{Mathematical background.}
\label{math-bg}
The underlying mathematical operations in HE are based on \emph{rings}.
A ring $\mathcal{R}$ is a set equipped with two binary operations: addition ($+$) and multiplication ($*$) which satisfy a set of axioms.
A \emph{polynomial ring} $\mathcal{R}[x]$ consists of polynomials as elements. 
For example, $\mathbb{Z}[x]$ represents the set of polynomials with integer coefficients.
We represent a polynomial $p(x) \in \mathcal{R}[x]$ of degree $n$ as:
\[p(x) = a_0 + a_1x + a_2x^2 + \ldots + a_nx^n = \sum_{k=0}^{n}a_kx^k\]
The elements $a_{i\in \{0,\ldots,n\}}$ are the coefficients of $p(x)$. 
The latter can thus be represented as a vector \textbf{a} $= [a_0, a_1, a_2, \ldots, a_n]$.
These coefficients can be bounded by applying a modulo $q$.\footnote{FHE operations introduce noise in ciphertexts; the ciphertext modulus acts as a bound to control noise growth.}
The associated ring is denoted $\mathcal{R}_q[x]$.
Further, a \emph{quotient ring} $\mathcal{R}_q[x]/\langle p(x)\rangle$ can be obtained by taking all polynomials in $\mathcal{R}_q[x]$ modulo the polynomial $p(x)$.
The security of HE is based on a hard computational problem called \emph{ring learning with errors} (RLWE)~\cite{rlwe}, which involves finding a hidden secret polynomial from a set of noisy polynomials in a ring structure. 
The most common choice for ciphertext spaces in FHE libraries is the ring $\mathbb{Z}_q[x]/\langle \Phi_n(x)\rangle$, where $\Phi_n(x)$ is a cyclotomic polynomial~\cite{shai2014,blanco2023fast}, \eg $x^n + 1$, and $n$ a power of 2.
Cyclotomic polynomials are used because their algebraic properties allow for efficient polynomial evaluations and guarantee the hardness of the RLWE problem.

Achieving a certain security level in HE, \eg 128-bit security, requires properly choosing the pair $(q,n)$.
The Homomorphic Encryption Standard~\cite{he-std} provides useful information regarding the choice of the parameters required to achieve a desired security level.
In practice, the value of $q$ is usually large, up to a thousand bits wide, which surpasses the capacity of a standard machine word of 64 bits. 
In such scenarios, techniques such as \emph{residual number systems} (RSN) are used to decompose large integers to fit into smaller integers.

\smallskip\noindent\textbf{Modular arithmetic units.} The arithmetics in homomorphic encryption schemes are performed over a finite field. 
The operations mainly include: coefficient-wise addition ($mod~q$), coefficient-wise multiplication ($mod~q$), and permutations~\cite{feldmann21}.
This means for a polynomial operation like addition, the operation over the coefficients is applied as shown in \autoref{eq:modulus}, which portrays a modular adder where $a$ and $b$ are coefficients of the polynomials.

\begin{equation} \label{eq:modulus}
    (a + b) \mod q = 
    \begin{cases} 
    a + b & \text{if } a + b < q \\
    a + b - q & \text{if } a + b \geq q 
    \end{cases}
 \end{equation}
These modular operations become very expensive, especially for modular multiplication with large polynomial coefficients.
To mitigate this problem, algorithms such as Barrett reduction~\cite{Barrett-reduction,Barrett} or Montgomery reduction~\cite{Montgomery} are employed to efficiently perform modular operations on large numbers without slow trial division. 
\section{Programming with UPMEM PIM}
\label{sec:arch}
A UPMEM PIM program typically consists of two parts: \emph{(1)} the host program, which is run by the CPU, and \emph{(2)} the program embedded and executed on the DPUs.
The host program must be implemented in C, C++, Java, or Python as of this writing due to the limited support of UPMEM's SDK~\cite{upemsdk}, and is used primarily for PIM orchestration, \ie initializing DPUs, distributing data across DPUs, launching the DPUs, and retrieving and aggregating partial results from the DPUs.
The DPU program, on the other hand, must be implemented in C for the same reason.
The UPMEM PIM SDK provides tools enabling developers to build and debug DPUs and host programs.

\subsection{Distributing data on DPUs}
To efficiently use PIM DPUs, we split the dataset (polynomials in this instance) into parts processed by the DPUs, similar to the approach used for GPU processing.
The processing operations per DPU are then shared among the DPU's tasklets. 
For example, consider a scenario where we aim to calculate the sum of the entries of a vector with $n$ elements, $V = [v_0, v_2, ..., v_{n-1}]$.
For simplicity, assume we can split the vector into $N_D$ \textbf{equally sized}\footnote{This mathematical modeling example is valid only for equally sized partitions, which we use in our experiments.} blocks for a system with $N$ DPUs.
Each DPU, $D_{i\in\{0,1,\ldots,N-1\}}$ processes a partition of the vector: $P_i = [v_{iN_D}, ..., v_{(i+1)N_D-1}]$.
The host CPU copies these partitions to the MRAM bank of the corresponding DPU, and in each DPU, the processing operations (summing vector elements) are shared among $T$ tasklets, with each tasklet processing a subset of the partition
in WRAM.

Similarly, assuming $D_i$'s partition ($P_i$) can be further \textbf{split evenly} among the $T$ tasklets, each tasklet $T_{ij}$ ($j^{th}$ tasklet of $i^{th}$ DPU) processes $N_T = \nicefrac{N_D}{T}$ elements which correspond to the elements of $v$ from index: $(iN_D + jN_T)$ to $iN_D + (j+1)N_T - 1$ inclusive, $i\in [0,N[$ and $j\in [0,T[$.
It is recommended to use at least 11 tasklets per DPU to fully utilize the DPU's 11-stage pipeline~\cite{gomez22}.
The individual results from the DPUs are then copied back from MRAM into the host's main memory, and aggregated to obtain the final result.
These copy operations must respect the size and alignment constraints imposed by the UPMEM DPU architecture. 
For example, all source and target addresses when copying data to MRAM or WRAM must be 8-byte aligned.
Algorithm~\autoref{algo:dpu_exec} summarizes this generic workflow for distributing data and processing operations to a set of DPUs.
\begin{algorithm}[!t]	
	\caption{DPU execution workflow}
	\begin{algorithmic}[1]
		\State $N$: number of DPUs
		\State $N_D$: data size per DPU
		\State $T$: number of tasklets per DPU
		\State $N_T$: data size per tasklet
		\State Host initializes DPUs
		\For{$i \gets 0$ \textbf{to} $(N-1)$}
			\State Copy data of size $N_D$ from host to DPU$_i$'s MRAM
			\State // Execute program on DPU$_i$				
			\For{$j \gets 0$ \textbf{to} $(T-1)$}
				\State Tasklet $T_{ij}$ copies chunk of size $N_T$ into WRAM
				\State $T_{ij}$ processes $N_T$ elements to obtain partial result
			\EndFor
			\State Aggregate partial results from tasklets
			\State Copy results from DPU$_i$ to host				
		\EndFor	
		\State Host aggregates results from all DPUs	
	\end{algorithmic}
	\label{algo:dpu_exec}
\end{algorithm}
\vspace*{-5pt} 

\observ{The programmer must determine how to effectively partition the data among the DPUs and tasklets.}

\subsection{Adapting existing libraries for UPMEM PIM DPUs}
All DPU code must be implemented in C.
This presents a serious constraint at the implementation language level, requiring developers to port existing (non C) code to C to use DPUs.
This can pose significant engineering challenges, especially for complex applications.
We focus on two open-source HE libraries, OpenFHE~\cite{openfhe22} and HElib~\cite{helib}, both implemented in C++. 
Given that UPMEM's SDK supports host applications in C++, we opt to reimplement only specific parts of these libraries in C to run on DPUs.
These parts consist of code to perform polynomial arithmetic.
Subsequently, we incorporate code in the host application or library to enable the distribution of input data (\eg polynomial coefficients/vectors) to the DPUs and retrieval of partial results from the DPUs.


\observ{Several HE libraries are implemented in C++. Developers are thus required to reimplement the target HE algorithms in C, which is the only language supported by UPMEM's DPU compiler toolchain at the time of this writing.}

\section{Evaluation}
\label{sec:eval}

The evaluation seeks to answer the following questions:

\begin{itemize}[leftmargin=28pt,nosep]
	\item[\textbf{Q1}:] How do PIM DPUs affect the cost of polynomial operations, \ie addition, multiplication, convolution, compared to CPU-based baselines?
	\item[\textbf{Q2}:] What are the main sources of overhead with DPU-based polynomial evaluations?
	\item[\textbf{Q3}:] How does the performance of polynomial evaluations scale with the number of DPUs?
\end{itemize}

We first address these questions by implementing polynomial addition and multiplication, comparing the results against multithreaded CPU baselines (\S\ref{ssec:poleval}).
Then, we modify the open-source libraries OpenFHE and HElib to leverage DPUs for these operations, and we assess the performance relative to the same libraries using CPU baselines (\S\ref{ssec:casestudy}).

\smallskip\noindent\textbf{Experimental setup.}
Our evaluation is conducted on a two-socket server.
Each socket is equipped with a 10-core Intel Xeon Silver 4210R processor clocked at 2.40GHz, with hyper-threading enabled.
Each processor has a 32KB L1d cache, 32KB L1i cache, 1MB L2 cache, and 13.75MB L3 cache.
The server comprises 32 ranks of DPUs, for a total of 2048 DPUs (each clocked at 350MHz) and 128GB of MRAM.
The memory bandwidth between each DPU and its associated MRAM bank is 1GB/s, representing an aggregate DPU memory bandwidth of up to 2.048TB/s on the server.
The machine runs Debian GNU/Linux 10 with the kernel version 4.19.0--26.

\smallskip\noindent\textbf{Methodology.}
We assess and compare DPU-based polynomial operations with their CPU-based counterparts.
The metrics evaluated include host CPU execution time (\cpu), total DPU execution time without data copying to or from the host (\dpu), host to DPU copy time (\hostdpu), and DPU to host copy time (\dpuhost).
Unless otherwise stated, the CPU-based experiments are run with 16 threads, and DPUs use 16 tasklets.
All experiments are run 100 times after 20 warmup iterations, and the median value is reported.

We first evaluate generic polynomial operations: addition and multiplication (convolutions and coefficient-wise) in the ring $\mathbb{Z}_q[x]/\langle x^n + 1\rangle$, with $q = 2^{16}+1$, on polynomials of varying sizes (or dimensions), \ie values of $n =$ number of polynomial coefficients.
The latter are randomly generated positive integers modulo $q$.
Then, we evaluate polynomial addition and multiplication operations using the APIs provided by OpenFHE and HElib, two popular HE libraries.

\subsection{Polynomial evaluations}\label{ssec:poleval}

As shown in related work~\cite{openfhe22}, the performance bottlenecks (both compute and memory) in FHE algorithms stem primarily from polynomial evaluations, \ie \emph{polynomial multiplication} and \emph{addition}. 
We investigate the usage of PIM DPUs to enhance these operations.

Consider two polynomials $p_1(x), p_2(x) \in \mathbb{Z}_q[x]/\langle x^n + 1\rangle$ with $p_1(x) = \sum_{k=0}^{n-1}a_kx^k$ and $p_2(x) = \sum_{k=0}^{n-1}b_kx^k$. 
In all our experiments, $n$ is a power of 2.

\begin{figure}[!t]
	\vspace{-4mm}
	\centering	
	\includegraphics[scale=1]{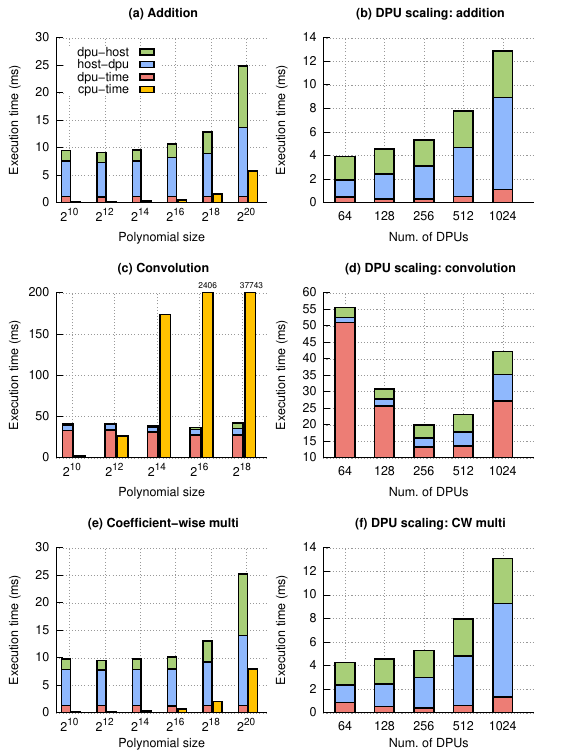}
	\caption{Cost of polynomial operations on CPU and DPUs.}
	\label{fig:poly-eval}
	\vspace{-6mm} 
\end{figure}
\smallskip\noindent\textbf{Polynomial addition.} The sum of both polynomials $p_1(x) + p_2(x)$ is given by:
$ p_1(x) + p_2(x) = \sum_{k=0}^{n-1}((a_k + b_k)~\text{mod}~q)x^k $

representing $n$ addition operations. 

\autoref{fig:poly-eval} (a) outlines the results obtained for performing polynomial additions on two polynomials of different sizes (\ie $n$).
For relatively smaller polynomials, \ie between $2^{10}$ and $2^{16}$, the CPU-based operations perform better, up to $7\times$ faster, when compared to the DPU-based variants.
However, for larger polynomials, \eg $2^{18}$ and $2^{20}$, the DPU versions perform better, up to $5\times$ faster relative to the CPU-based variants, when considering only execution times.

\observ{
DPU-based computations perform better with respect to the CPU-based version for larger datasets. This is due to the higher parallelism offered and the improved memory bandwidth of DPUs with respect to DRAM bandwidth. 
}
However, the overhead of copy operations between the host and DPUs (\hostdpu and \dpuhost) dominates the overall cost of the DPU operations. 
Copy operations from the host to the DPUs are more costly compared to those from the DPUs back to the host.
This is because 2 polynomials of degree $n$ ($2n$ coefficients) are copied from host to the DPUs vs $1$ polynomial of degree $n$ (the result of polynomial addition) copied from the DPUs to the host.
The large cost of copy operations is consistent with prior work~\cite{hyun2024, gomez22}.

\observ{
	Copying data from the host to DPUs and vice versa is the dominant cost for most DPU-related operations.
}

\smallskip\noindent\textbf{Polynomial multiplication.} A naive (textbook) multiplication of both polynomials can be obtained as follows:

\[p_1(x)*p_2(x) = \sum_{k=0}^{2n-2}c_kx^k, ~~\text{where}~~c_k = \sum_{i+j=k}^{}a_ib_j~\text{mod}~q\]

The vector $\boldsymbol{c} = [c_0, c_1, \ldots, c_{2n-2}]$ is referred to as the \textit{convolution} of the vectors $\boldsymbol{a} = [a_0, a_1, \ldots, a_{n-1}]$ and\linebreak $\boldsymbol{b} = [b_0, b_1, \ldots, b_{n-1}]$.
Computing the convolution as such involves multiplying the coefficients of the first polynomial by every other coefficient of the second polynomial, necessitating a total of $n^2$ multiplication and $(n-1)^2$ addition operations.
This results in a computational complexity of $\mathcal{O}(n^2)$, making it unsuitable for larger degree polynomials.

%

\autoref{fig:poly-eval} (c) outlines the cost of performing convolutions (\ie naive polynomial multiplication) on DPUs as compared to a CPU-based approach. 
For relatively smaller polynomials (number of coefficients = $2^{10}, 2^{12}$), the CPU-based variants perform better, up to 3.7$\times$ faster compared to the DPU-based variants.
However, for larger polynomials (number of coefficients = $2^{16}, 2^{18}$), the DPU-based variants execute much faster than the CPU-based operations, up to 1397$\times$ and 898$\times$ faster without and with DPU copy operations taken into account respectively.
While this may initially appear counterintuitive since DPUs are much slower than the host CPU, it is justified by the fact that a large number of DPUs, \eg 1024, operating in parallel at 350MHz, can collectively execute more instructions per second than a single CPU running at 2.5GHz.
The advantage of DPU parallelism is particularly significant in textbook convolution implementations because the latter involves a much larger number of operations (addition and multiplication) when compared to coefficient-wise addition or multiplication for polynomials of similar size.
Additionally, the large number of operations involved amortizes the cost of data copy operations between the host CPU and the DPU.

\observ{Performing extensive operations on DPUs can offset the overhead of data copying between the host CPU and DPUs, resulting in improved performance on DPUs.}

A more efficient approach for performing polynomial multiplication involves using the \emph{discrete fourier transform} (DFT) of the polynomials. 
The DFT algorithm transform polynomials into the frequency domain, allowing convolutions ($*$) to be computed by performing coefficient-wise multiplications ($\odot$) on the fourier transforms of the vectors and applying the inverse fourier transform on the result, as shown in the following equation:
\begin{equation}
	\mathcal{F}(a*b) = \mathcal{F}(a)\odot \mathcal{F}(b) \Longrightarrow a*b = \mathcal{F}^{-1}(\mathcal{F}(a)\odot \mathcal{F}(b))
	\label{eqn:dft}
\end{equation}
where $\mathcal{F}$ and $\mathcal{F}^{-1}$ represent the discrete fourier transform and its inverse respectively. \autoref{eqn:dft} is commonly referred to as the \emph{convolution theorem}.
A popular algorithm for evaluating the DFT is the \emph{number theoretic transform} (NTT)\cite{NTT1,NTT2,NTT3}, which is a particular case of the DFT over the ring $\mathbb{Z}_q[x]$.
Given a vector of $n$ elements $\boldsymbol{a}$, NTT transforms $\boldsymbol{a}$ into another $n$ element vector $\boldsymbol{\hat{a}}$. That is, $\boldsymbol{\hat{a}} = NTT(\boldsymbol{a})$ such that:
\[
\hat{a}_j = \sum_{i=0}^{n-1}\omega^{ij}a_i~\text{mod}~q,~j \in [0, n[
\]

where $\omega$ represents the primitive $n$-th root of unity. The latter is a number in $\mathbb{Z}_q$ such that $\omega^n = 1~\text{mod}~q$ holds.
Similarly, the inverse NTT transform (INTT) can be applied to $\boldsymbol{\hat{a}}$ to obtain
$\boldsymbol{a}$. That is, $\boldsymbol{a} = INTT(\boldsymbol{\hat{a}})$ such that:
\[
a_i = \frac{1}{n}\sum_{j=0}^{n-1}\omega^{-ij}\hat{a}_j~\text{mod}~q,~i \in [0, n[
\] 
The NTT algorithm employs the convolution theorem outlined in \autoref{eqn:dft} to compute the convolution $\boldsymbol{c}$ of $\boldsymbol{a}$ and $\boldsymbol{b}$ via coefficient-wise multiplications as follows:
\[
\boldsymbol{c} = INTT(NTT(\boldsymbol{a})\odot NTT(\boldsymbol{b}))
\]
This is referred to as the \emph{cyclic convolution} or \emph{positive wrapped convolution} (PWC)~\cite{NTT2}, and is commonly used to compute polynomial multiplications in $\mathbb{Z}_q[x]/\langle x^n - 1\rangle$.
Fast fourier transform (FFT) algorithms~\cite{cooley1965algorithm,gentleman1966fast} are leveraged to achieve an improved computational complexity of $\mathcal{O}(n\log n)$ for computing convolutions via NTT.
We evaluate the cost of performing coefficient-wise multiplications, an important part of NTT, on DPUs. 
\autoref{fig:poly-eval} (e) outlines the results obtained.


Similar to the preceding experiments, we observe better performance for the CPU-based operations for smaller polynomial sizes.
More specifically, for polynomials of size between $2^{10}$ and $2^{16}$, the CPU-based variants are up to 9$\times$ faster when compared to the DPU-based variants when DPU data copy operations are not taken into account.
For larger polynomials, \ie of size $2^{18}$ and $2^{20}$, the DPU-based operations are up to 6$\times$ faster than the CPU baselines when copy operations are not taken into account.
The DPUs leverage improved parallelism to achieve faster execution times here.
On the other hand, when DPU copy operations are taken into account, the CPU-based performs better, executing up to 3$\times$ faster relative to the DPU-based variants.
This poorer performance of the DPU-based variants relative to the CPU baselines is due to the small ratio of on-DPU operations (multiplications) to the size of data copied, thereby overshadowing the benefits of DPU parallelism.

In general, coefficient-wise multiplications on the DPU are more expensive when compared to coefficient-wise addition operations.
As explained in~\cite{gomez22}, this is due to the absence of a $32\times 32$ bit multiplier in the DPU pipeline as opposed to addition which has native support.
\observ{There is potential for DPUs to accelerate algorithms like the number theoretic transform if the cost of data copying between the host and DPUs is reduced.		}

\renewcommand{\arraystretch}{1.75}
\setlength{\tabcolsep}{2.3pt}
\begin{table}
	\centering
	\rowcolors{3}{gray!0}{gray!10}
	\begin{tabularx}{\columnwidth}{lcccc} 
		\toprule
		\textbf{Operation} & \textbf{host$\rightarrow$DPU} & \textbf{DPU$\rightarrow$host} & \textbf{Num. of ops} & \textbf{$\alpha$} \\
		\midrule
		Addition & $2n$ & $n$ & $n$ & $\frac{1}{3}$ \\
		CW multiplication & $2n$ & $n$ & $n$ & $\frac{1}{3}$\\
		Convolution & $2n$ & $n$ & $n^2 + (n-1)^2$ & $\frac{n^2 + (n-1)^2}{3n}$ \\
		
		\bottomrule
	\end{tabularx}
	\caption{Ratio ($\alpha$) of on-DPU operations to size of copied data for different polynomial operations. For example, in polynomial addition, two polynomials with $n$ coefficients each are copied from the host to the DPUs (\ie host$\rightarrow$DPU $= 2n$), coefficient-wise addition operations (\ie $n$ in total) are performed on the DPUs, and the resulting polynomial with $n$ coefficients is copied from the DPUs to the host (\ie DPU$\rightarrow$host $=n$). Total data copied in both directions $=2n + n = 3n$ and total number of on-DPU operations $=n$, hence $\alpha = \nicefrac{n}{3n} = \nicefrac{1}{3}$.  Our results indicate better overall performance on PIM for larger values of $\alpha$.}
	\label{tab:ratios}
\end{table}
\autoref{tab:ratios} represents the ratios ($\alpha$) of DPU operations (additions or multiplications) to the size of data copied for different polynomial operations.
In general, for small values of $\alpha$, \ie $\alpha < 1$, we observe relatively poor performance for PIM variants because data copies dominate the cost.
For larger values of $\alpha$, \eg $\nicefrac{n^2 + (n-1)^2}{3n}$ (for convolutions) PIM provides better performance compared to the CPU-based operations, up to 898$\times$ in the case of $n = 2^{18}$.


\smallskip\noindent\textbf{DPU scaling.}
\autoref{fig:poly-eval} (b), (d), and (f) represent the performance of the corresponding polynomial operations for polynomials of size $2^{20}$ when the operations are parallelized across 64, 128, 256, 512, and 1024 DPUs.
Overall, we observe improved performance for DPU execution times as the number of DPUs increases up to 512 DPUs.

\observ{In general, the performance of DPU-based operations scales with an increasing DPU count.}

Yet, with a larger DPU count, \ie 1024, we observe an increase in DPU execution times, likely stemming from increased overhead associated with launching a larger number of DPUs.

\observ{The overhead associated with launching DPUs can lead to longer overall execution times for a larger number of DPUs compared to a smaller number.}

In spite of smaller data chunks per DPU with an increasing number of DPUs, the host-to-DPU data copy times increase with the number of DPUs. 
This can be attributed to the increased number of copy operations as a result of increasing the number of DPUs and suggests some inefficiency in the rank transfer functions \code{dpu\_prepare\_xfer} and \code{dpu\_push\_xfer} provided by the UPMEM API used to push different buffers to a set of DPUs in one transfer.


\subsection{Case studies with existing HE libraries}\label{ssec:casestudy}
We follow up with a case study analysis of UPMEM PIM DPUs on the performance of two popular HE libraries: OpenFHE and HElib. 
These libraries are originally implemented in C++.
As of this writing, the UPMEM compilation toolchain supports only C-based DPU code. 
Therefore, we adapted the HE libraries accordingly to support DPUs.

\observ{Many popular HE libraries are implemented in C++. Developers are thus required to reimplement the target HE algorithms in C, which is currently the only language supported by UPMEM's DPU compiler toolchain.}

\smallskip\noindent\textbf{OpenFHE.}
This library~\cite{openfhe22} is an open-source software that supports fully homomorphic encryption.
It implements homomorphic encryption schemes like BFV, BGV, CKKS, Ducas-Micciancio (FHEW), and Chillotti-Gama-Georgieva-Izabachene (TFHE).
Furthermore, it provides FHE extensions such as Proxy ReEncryption (PRE) and Threshold FHE.
OpenFHE is developed around the principles of modularity, extensibility, and efficiency providing ease of application deployment across multiple FHE schemes and hardware accelerators.
Specifically, the library includes a hardware acceleration layer (HAL), allowing a range of hardware accelerators to integrate seamlessly.
Finally, on an efficiency basis, for BGV-like schemes (BFV, BGV and CKKS), only the RNS (Residual Number System) variants of the schemes are supported.

\smallskip\noindent\textbf{RNS in OpenFHE.} 
The ciphertext modulus $q$ is decomposed into $k$ smaller $q_i$ (subdomains) such that $q = \prod_{i=1}^{k}q_i$, where each $q_i$ is a prime number that fits in a standard machine word. 
Every $q_i$ can be processed in parallel during the modular operations.
From now on, we refer to each subdomain associated with the modulus $q_i$ as a \emph{tower}. 
In essence, a tower represents a vector or a polynomial whose coefficients are sampled to the modulo $q_i$.
To accelerate polynomial operations, polynomials in OpenFHE are represented following the \emph{Double Chinese Remainder Theorem} (DCRT) representation~\cite{mono2023finding}. 
 
\smallskip\noindent\textbf{Evaluations}. 
We evaluate homomorphic additions and multiplications for ciphertexts encoded as polynomials in DCRT form.
As explained previously, HE libraries like OpenFHE implement homomorphic algorithms based on polynomial arithmetic.
In integrating the UPMEM PIM with OpenFHE, we utilize OpenFHE's HAL, whose primary objective is to accommodate multiple hardware acceleration backends for bottleneck polynomial and RNS operations. 
This implies a distinct implementation for integers, modular vectors, NTT transformations, and polynomials, utilizing the same cryptographic capabilities across all backends.

The CPU implementation of polynomial arithmetic is measured under both single-threaded and multithreaded configurations by toggling the respective compiler flags for OpenMP. 
The multithreaded implementation of polynomial arithmetic involves executing the operation on the respective RNS-subdomain polynomial vectors. 
Specifically, the number of threads is determined based on the available number of towers.
For our PIM-based implementations, the primary objective focuses on partitioning the data to be transferred to the DPUs. 
This is accomplished by dedicating a specific number of DPUs per tower.
We leverage a total of $2048$ DPUs for evaluations on OpenFHE. 
For polynomial addition, the same modular adders as described in \S\ref{math-bg} are utilized, while for polynomial multiplication, the Barrett reduction\cite{Barrett,Barrett-reduction} method is employed.

\begin{figure}[!t]
	\centering	
	\includegraphics[scale=1]{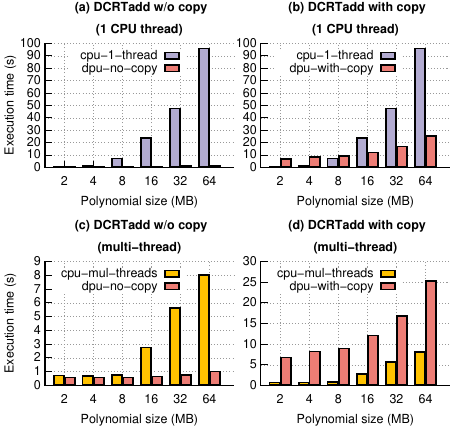}
	\caption{Cost of polynomial addition on DPUs in OpenFHE with and without the host to DPU data copying vs. single and multi-threaded CPU variants.}
	\label{fig:openfhe-add}	
\end{figure}

\smallskip\noindent\textbf{OpenFHE polynomial addition.}
\autoref{fig:openfhe-add} outlines the results of polynomial addition operations for polynomials of varying sizes on DPUs and the host CPU (single and multi-threaded).
\autoref{fig:openfhe-add} (a) and (c) do not take into consideration the cost of copying data between the host and the DPUs, while \autoref{fig:openfhe-add} (b) and (d) include the cost of data copying.
When the cost of data copying is not taken into consideration, we observe much better performance for the PIM-based variant, which performs up to 97$\times$ and 8$\times$ faster when compared to the single and multi-threaded CPU-based variants, respectively.
This is mainly a result of the massive parallelism offered by UPMEM's PIM architecture.
However, the multithreaded CPU baseline performs better (up to $3\times$ faster) when DPU copy operations are included in the total cost.
Even though individual DPUs are slower than the host CPU, the large degree of parallelism offered by the PIM architecture provides better performance for the PIM-based variants with respect to the single-threaded CPU variant, even when copy operations are taken into account.

\begin{figure}[!t]
	\centering	
	\includegraphics[scale=1]{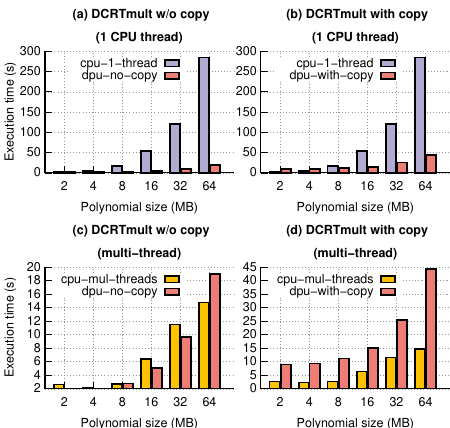}
	\vspace{1.5mm}
	\caption{Cost of polynomial multiplication on DPUs in OpenFHE with and without host to DPU data copying vs single and multi-threaded CPU variants.}
	\label{fig:openfhe-multi}
	\vspace{-3mm}	
\end{figure}

\smallskip\noindent\textbf{OpenFHE polynomial multiplication.}
\autoref{fig:openfhe-multi} outlines the results of polynomial multiplication operations for polynomials of varying sizes on DPUs and the host CPU (single and multi-threaded).
\autoref{fig:openfhe-multi} (a) and (c) do not take into consideration the cost of copying data between the host and the DPUs, while \autoref{fig:openfhe-multi} (b) and (d) include the cost of data copying.
Similar to OpenFHE polynomial addition, when compared to the single-threaded CPU baseline, the DPU variants perform better, up to 15$\times$ and 6$\times$ without and with data copying taken into consideration, respectively.
As previously explained, this is due to the improved parallelism provided by the PIM architecture.
Conversely, when data copy costs are taken into account, the DPU variants perform poorly, with copy operations dominating the overhead.

\smallskip\noindent\textbf{NTT in OpenFHE.}
As previously explained in \S\ref{ssec:poleval}, the number theoretic transform can be used to compute the convolution $\boldsymbol{c}$ of two vectors $\boldsymbol{a}$ and $\boldsymbol{b}$ using coefficient-wise multiplication.
To mitigate the overhead of zero padding vectors required in computing the PWC over $\mathbb{Z}_q[x]/\langle x^n + 1\rangle$, the \emph{negacyclic} or \emph{negative wrapped convolution} (NWC)~\cite{NTT2} can be employed to perform NTT.
It uses the primitive $2n$-th root of unity, denoted as $\psi$ where $\psi^2 = \omega~\text{mod}~q$ and $\psi^n = -1~\text{mod}~q$. 
We denote NTT based on the NWC as NTT$^\psi$. 
Given a vector of $n$ elements $\boldsymbol{a}$, NTT$^\psi$ transforms $\boldsymbol{a}$ into another $n$ element vector $\boldsymbol{\hat{a}}$. That is, $\boldsymbol{\hat{a}} = NTT^\psi(\boldsymbol{a})$ such that:

\[
\hat{a}_j = \sum_{i=0}^{n-1}\psi^{2ij + i}a_i~\text{mod}~q,~j \in [0, n[
\]

\noindent Similarly, the inverse NTT transform (INTT$^{\psi^{-1}}$) can be applied to $\boldsymbol{\hat{a}}$ to obtain
$\boldsymbol{a}$. That is, $\boldsymbol{a} = INTT^{\psi^{-1}}(\boldsymbol{\hat{a}})$ such that:
\[
a_i = \frac{1}{n}\sum_{j=0}^{n-1}\psi^{-(2ij + i)}\hat{a}_j~\text{mod}~q,~i \in [0, n[
\] 

OpenFHE implements NTT based on the NWC and leverages two popular FFT algorithms: Cooley-Tukey (CT) butterfly algorithm~\cite{cooley1965algorithm} for computing NTT$^\psi$ and the Gentleman-Sande (GS) butterfly algorithm~\cite{gentleman1966fast} for computing INTT$^{\psi^{-1}}$.
Both algorithms aim to reduce complexity and speed up matrix multiplication operations involved in NTT transformations. 
They follow a divide-and-conquer approach, splitting the computation of NTT$^\psi$ and INTT$^{\psi^{-1}}$ into two $\nicefrac{n}{2}$ parts. 
\begin{figure}[!t]
	\centering	
	\includegraphics[scale=0.34]{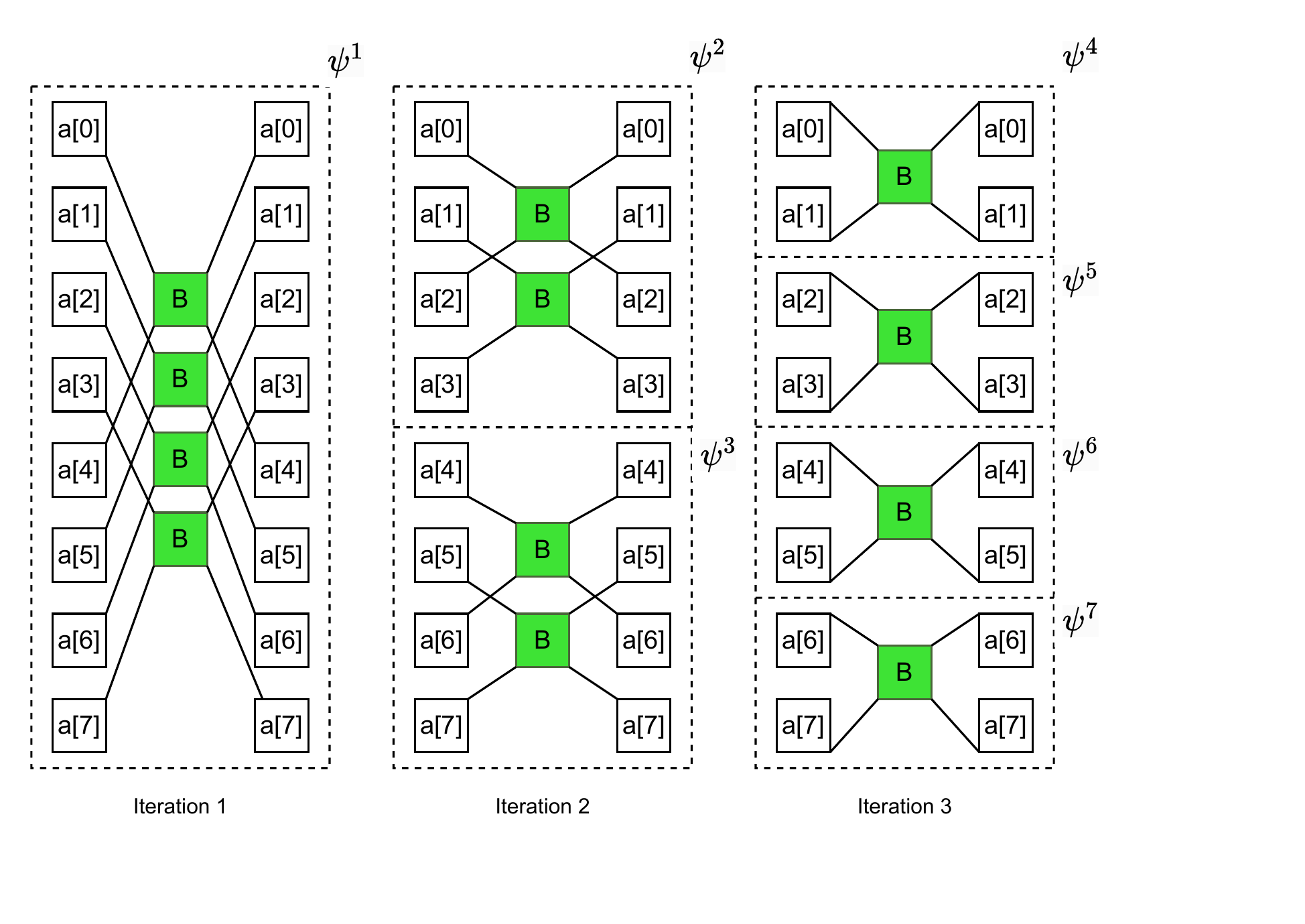}
	\vspace{-10mm}
	\caption{Progressive iterations of Cooley-Tukey butterfly operations in NTT for a polynomial of size 8. 
		The butterfly operation "B" entails modular addition and multiplication operations involving polynomial coefficients ($a[i]$) and associated powers of $\psi$ used in the computation.
		During each B operation, data is read from and written back to the same vector/array, \ie in-place computation.
		As a result, outputs of B operations from one iteration serve as inputs for other B operations in the next iteration, showcasing the data dependencies between iterations.}
	\label{fig:butterfly}
	\vspace{-4mm}	
\end{figure}
\autoref{fig:butterfly} illustrates how the Cooley-Tukey (CT) butterfly algorithm performs data accesses for each butterfly operation; a similar principle also applies to the Gentleman-Sande (GS) butterfly algorithm.
Butterfly (B) operations involve modular multiplication and addition operations of polynomial coefficients with factors of the $n$-th root of unity.
Our DPU-based implementation offloads these B operations to DPUs.
However, this presents a challenge due to the data dependencies between CT iterations, as the results of a B operation in one iteration are used as input to another B operation in the next iteration.
At the implementation level, this requires expensive copy operations between the DPUs and the host, as well as complex synchronization between DPUs via the host application.

\begin{figure}[!t]
	\centering	
	\includegraphics[scale=1]{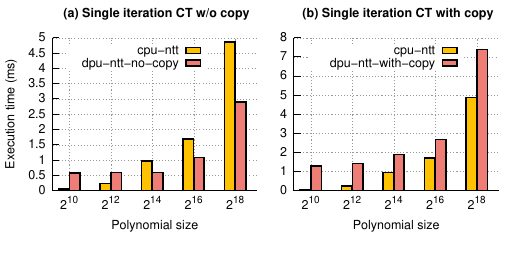}
	\caption{Cost of one iteration of OpenFHE Cooley-Tukey butterfly algorithm on DPUs with and without the host to DPU data copying vs. a single-threaded CPU baseline.}
	\label{fig:ct}	
\end{figure}
\autoref{fig:ct} outlines the results obtained for a DPU-based variant of the CT algorithm for polynomials of varying dimensions compared to a single-threaded CPU baseline. With data copying taken into account, the DPU variant is up to 21$\times$ slower when compared to the CPU baseline.
For a polynomial of size $2^n$, the total number of iterations required for the full NTT operation is $n$. 
This translates to larger overhead due to data copy operations and, thus, poorer performance for the DPU-based variant when compared to the CPU baseline.

\observ{In NTT algorithms like Cooley-Tukey and Gentleman-Sande, the primary bottleneck to effective DPU-based parallelism lies in the data copy operations between the host and the DPUs.}


\smallskip\noindent\textbf{HElib.} 
This library is an open-source C++ software that implements BGV and CKKS HE schemes.
We focus on its BGV scheme, which encodes ciphertexts in the polynomial ring $\mathcal{R}_q = \mathbb{Z}_q[x]/\langle x^n + 1\rangle$.
HElib's polynomial arithmetic is tightly coupled to a third-party library, NTL~\cite{ntl}.
As such, adapting it for PIM requires this library to be equally adapted for PIM, further increasing the complexity of the porting operation.
For this reason, we re-implemented some of NTL's polynomial algorithms directly in HElib and present results only for polynomial addition, used by HElib's BGV scheme.

\smallskip\noindent\textbf{Evaluations.}
We evaluate addition operations for polynomials in DCRT form.
The DPU-based variants are compared against single-threaded CPU-based variants.
\begin{figure}[!t]
	\centering	
	\includegraphics[scale=1]{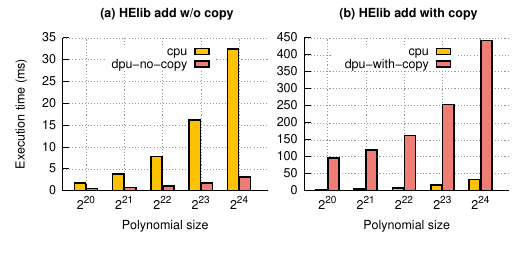}
	\caption{Cost of polynomial addition on DPUs in HElib with and without the host to DPU data copying vs. single-threaded CPU baseline.}
	\label{fig:helib-add}
	\vspace{-6mm}	
\end{figure}
\autoref{fig:helib-add} outlines the results obtained.
When copy operations are not taken into account, the PIM variant performs better, up to 10$\times$ faster, when compared to the CPU-based implementation.
However, in line with our prior observations (\ie OpenFHE), when copy operations are taken into account, the overall processing time on PIM is much larger, with the copy operations dominating the cost.
Consequently, the CPU-based implementation performs better, up to 13$\times$ faster. 
\section{Discussion}
\label{sec:discussion}
%
\smallskip\noindent\textbf{DPU copy operations.}
UPMEM's PIM architecture presents benefits in terms of enhanced parallelism and improved memory bandwidths.
These benefits are particularly evident in two situations: \emph{(1)} when dealing with extensive computations (\ie instructions executed), and \emph{(2)} when handling very large datasets, typically surpassing the CPU's last-level cache, which necessitates costly CPU-DRAM data movements.
DPUs address this by bringing computation closer to memory, thereby potentially mitigating these issues.
However, the transfer of data between the host DRAM and DPU memory remains a significant challenge.
Our findings indicate that the cost of these data movements tends to dominate DPU-related overhead, especially when relatively fewer operations are executed on the DPU per unit of data copied.
This observation is in line with recent work~\cite{lee2024}, which analyzes the data movement bottleneck in PIM architectures, as well as~\cite{nider21} which evaluates UPMEM PIM on commonly used workloads.

To achieve better overall performance (in terms of execution times) with DPUs, the benefits of increased parallelism and improved memory bandwidths should outweigh the losses due to data movements.
Our results suggest the cost of the data movements can be offset when a large number of computations are performed on the DPUs per unit of data copied.
To minimize data movements, DPU programs should be structured to perform copy operations only once in each direction: when distributing data to the DPUs (\hostdpu) and when transferring results back to the host (\dpuhost).

\smallskip\noindent\textbf{Zero-copy approach for DPUs.}
Based on our current findings, it is appropriate to conclude that a zero-copy approach for DPUs is the optimal scenario.
This entails storing the data directly in DPU memory, \ie MRAM, removing the need for MRAM-DRAM data copy operations.
This would be useful for use cases such as network applications, where TCP packet data is stored directly in PIM by the kernel and PIM-based processing leveraged for improved parallelism and memory bandwidths.
A zero-copy approach will also prove useful for use cases with application-generated data.

\smallskip\noindent\textbf{Code adaptation.}
As of this writing, UPMEM provides an SDK that only supports writing DPU-based programs in C and is expected to provide support for Rust in the near future.
This limitation at the language level can deter the adoption of UPMEM's PIM technology due to the need for modifications to existing codebases.
For instance, we encountered this constraint when integrating PIM in OpenFHE and HElib, which are written in C++.
This challenge can be addressed by expanding support of UPMEM PIM to include more high-level programming languages, either through native support or language bindings, or leveraging a common compilation target such as WebAssembly (Wasm)~\cite{DBLP:conf/pldi/HaasRSTHGWZB17}.
In the context of PIM, Wasm can serve as an abstraction to the hardware accelerator to offload parallel computations, thereby easing the software development and reasoning about complex code.
This approach also promotes decoupling PIM from the specific implementation of UPMEM, enhancing the portability of PIM-related code across heterogeneous hardware platforms~\cite{DBLP:conf/hpdc/MenetreyPFS22}.
Furthermore, code annotations~\cite{glamdring, secv} or attributes (some of which are already available in UPMEM's SDK, \eg \code{\_\_mram}, \code{\_\_dma\_aligned}) can be introduced or extended to mark PIM data to be copied to or from the DPU. 
The compiler can then use these annotations to optimize data transfer routines.
These solutions allow developers to use familiar programming languages and tools, reducing entry barriers and speeding up the integration of UPMEM's PIM technology into various application ecosystems.

\section{Related work}
\label{sec:rw}

We classify the related work into three categories:
\emph{(i)}~systems that leverage PIM architectures for accelerating homomorphic encryption,
\emph{(ii)}~systems that leverage PIM architectures for accelerating generic applications, and
\emph{(iii)}~systems that leverage other hardware architectures for accelerating homomorphic encryption.

\looseness-1
\smallskip\noindent\textbf{PIM-based acceleration for HE.}
Gupta \textit{et al.}~\cite{saransh21} conducted a theoretical analysis of the performance impact of PIM on enhancing polynomial addition and multiplication operations in HE and compared the results with the open-source library Palisade~\cite{palisade}, which has since been integrated into OpenFHE.
Building on this foundational work, our paper extends the research by benchmarking PIM using actual hardware and integrating HE operations into OpenFHE.
More recently, Gupta \textit{et al.}~\cite{gupta23} explored the performance of PIM using UPMEM hardware, with a specific focus on the homomorphic addition and multiplication operations under statistical workloads.
Their results revealed concrete performance impact of PIM as compared to CPU and GPU setups.
Our study addresses key aspects not covered in their work: we evaluate the cost of data movements between the host and DPUs for the homomorphic operations and employ multithreaded CPU baselines, as opposed to \cite{gupta23}, which limits comparisons to a single-core CPU baseline.
Finally, MemFHE~\cite{saransh24}, CryptoPIM~\cite{DBLP:conf/dac/NejatollahiGIRC20} and Jonatan \textit{et al.}~\cite{gilbert24} evaluated the impact of PIM on the NTT algorithm, commonly used in FHE libraries for efficient polynomial multiplication.
On the other hand, CiM-HE~\cite{DBLP:journals/tvlsi/ReisTJNH20} focused on BFV, another foundational scheme in FHE libraries.

Our work integrates PIM support into the lower-level polynomial arithmetic layer of OpenFHE and HElib libraries, thus preserving the existing API and ensuring that end-users continue to work with the same interface without the additional complexity introduced by PIM.

\smallskip\noindent\textbf{PIM-based acceleration for generic applications.}
A 2021 case study~\cite{nider21} evaluates the impact of PIM across five commonly used applications---compression, hyperdimensional computing, encryption, text search, and filtering---emphasizing the strengths of the UPMEM PIM architecture.
Besides, this architecture has been adopted by many other studies.
For instance, SparseP~\cite{Giannoula2022SparsePPomacs} offered an open-source initiative providing sparse matrix-vector multiplication, while D. Lavenier~\cite{DBLP:conf/bibm/LavenierRF16} explored applications in DNA sequencing.
PID-Join~\cite{DBLP:journals/pacmmod/LimLCLPKLK23} enhanced in-memory join algorithms for database systems, and PIM-STM~\cite{pimstm} developed an efficient software abstraction for transactional memory.
Additionally, Nider \textit{et al.}~\cite{DBLP:conf/systor/NiderDGNF22} focused on decoding images to produce high-quality outputs.

Aligned with these studies, our research further eases the use of UPMEM PIM architecture and introduces open-source implementations to promote further investigation and reproducibility in this field.

\smallskip\noindent\textbf{Non PIM-based hardware acceleration for HE.}
Over the past decades, numerous studies have accelerated HE by leveraging the versatility of FPGAs.
For example, F1~~\cite{feldmann21} proposes a programmable FHE accelerator, deeply specialized for FHE primitives such as modular arithmetic, NTT, and structured permutations.
Similarly, Sujoy \textit{et al.}.~\cite{sinha19} and HEAWS~\cite{DBLP:journals/tc/TuranRV20} focused on accelerating BFV multiplications, while HEAX~\cite{DBLP:conf/asplos/RiaziLPD20} enhanced CKKS multiplications.
These FPGA-based solutions face challenges with high data movement, a limitation potentially overcome by optimal PIM implementations.
Furthermore, Intel has improved single-threaded HE computations using their advanced vector extensions (Intel AVX512)~\cite{DBLP:conf/ccs/BoemerKSSG21}, which are now part of the open-source libraries OpenFHE and SEAL~\cite{sealcrypto}.
Our practical report, in contrast, emphasizes the benefits of parallel execution enabled by DPUs.
Prior work has explored the optimization of HE computations using GPUs, exploiting the highly parallel architecture of graphics cards to speed up FHE operations~\cite{wang12, DBLP:conf/iscas/WangCH14, DBLP:journals/tc/WangHCHS15, DBLP:journals/tetc/BadawiPAVR21, shivdikar23}.
While these GPU-based approaches generally surpass traditional CPU methods, they still require the transfer of data from main memory to dedicated GPU memory, a drawback that PIM technology aims to address.

\section{Conclusion}
\label{sec:conclusion}
In this work, we investigate the potential for enhancing homomorphic encryption operations with UPMEM's PIM hardware.
We evaluate key polynomial operations such as addition and multiplication, which underpin HE algorithms, employing both generic implementations as well as specific implementations from two popular HE libraries, OpenFHE and HElib, and we provide valuable insights based on our findings.
In summary, our findings indicate good performance for DPU execution on large datasets, attributed to the extensive parallelism provided by PIM hardware.
Unlike prior studies that overlook the overhead due to data copying between the host and DPU address spaces, our evaluations consider this significant overhead and show that it dominates DPU-related costs.
This overhead is responsible for the overall poorer performance of PIM-based systems compared to their CPU-based counterparts.
Our results underscore the need to optimize data movement strategies to fully leverage PIM hardware's capabilities.
The performance gap between PIM-based and CPU-based operations could be narrowed by minimizing the overhead associated with data transfers between the host CPU and the DPUs.
Future research efforts could focus on developing efficient data transfer mechanisms tailored to the unique architecture of PIM systems. 


\section*{Acknowledgment}
This work was supported by the Swiss National Science Foundation under project P4: Practical Privacy-Preserving Processing (no. 215216).

\bibliographystyle{plain}
\bibliography{biblio}

\end{document}